\documentstyle[psfig,twocolumn,aps]{revtex}
\begin{document}
\title{Transverse mode coupling in a Kerr medium}
\author{Jean-Michel Courty$^a$ and Astrid Lambrecht$^b$}
\address{$(a)$ Laboratoire Kastler Brossel, UPMC, ENS, CNRS\\
4 place Jussieu, F-75252 Paris Cedex 05, France \\
$(b)$ Max-Planck-Institut f\"ur Quantenoptik\\
Hans-Kopfermann-Str.1, D-85748 Garching, Germany}
\date{{\sc Physical Review} {\bf A 54}, p.5243 (1996)}
\maketitle

\begin{abstract}
We analyze non-linear transverse mode coupling in a Kerr medium placed in an
optical cavity and its influence on bistability and different kinds of
quantum noise reduction. Even for an input beam which is perfectly matched
to a cavity mode, the non-linear coupling produces an excess noise in the
fluctuations of the output beam. Intensity squeezing seems to be
particularly robust with respect to mode coupling while quadrature squeezing
is more sensitive. However it is possible to find a mode the quadrature
squeezing of which is not affected by the coupling.\\[2mm]
PACS number(s): 42.65.Sf 
\end{abstract}

%EndAName

\section{Introduction}

During the last years different kinds of systems for the generation of 
squeezed light have been proposed and realized. The bistable device obtained 
by placing a non-linear
Kerr medium inside an optical cavity is one of them. When the bistability
turning point is approached,  the fluctuations in one of the field
quadratures of the incoming beam are reduced below shot noise \cite
{Luigi1,Xiao1,Reid,Hilico}.  Nearly perfect squeezing is expected when the
non-linear medium is a lossless Kerr medium \cite{Collett,Shelby,Reynaud1}.
Squeezing via the optical Kerr effect was experimentally demonstrated at the
output of a bistable cavity containing an atomic beam\cite{Raizen,Hope1} or
a cloud of cold atoms released from a magneto-optical trap \cite{Lambrecht}.

Most theoretical treatments have simplified the description of the
field-matter interaction by modeling the laser beam as a plane-wave [1-7].
Thanks to these works, the squeezing mechanism in this system is well
understood and it has been shown that this kind of device has the
potential to generate very large squeezing. There is now a need for more
complete models, either to analyze experimental results of current
experiments or to estimate the limitations of future set-ups. For more
realistic systems it is clear that the plane wave description has to be
modified to account for the spatial structure of the laser beam. When
dealing with a non-linear medium made of atoms, a first step consists in
averaging the mean field as well as the fluctuations over the beam profile
in order to take into account the differences of light intensity and of
coupling strength at different localizations inside the beam \cite
{Drummond,Xiao2,Hope2,Lambrecht2}. This leads to a modification of the
bistability curve and of the squeezing obtainable with the device. Only in
the limit of an ideal Kerr medium (characterized by a $\chi ^{\left(
3\right) }$ susceptibility) this single transverse mode approach leads to
results identical to the plane wave model \cite{Lambrecht2}.

It is also well known that the interaction of a light beam with a non-linear
medium gives rise to transverse spatial effects which may have a strong
influence on the quantum fluctuations of the beam \cite{Luigi2,Luigi3}. For
a Kerr medium, the first effects of the non-linearity (before the appearance
of spatial instabilities) are self focusing or self defocusing. When dealing
with a medium inside an optical cavity, these effects produce a coupling
between different transverse modes, which depends on the intracavity
intensity and changes not only the mean field but also field fluctuations.
In particular the non-linear coupling produces an excess noise which is
appreciable even if the cavity is perfectly matched to the spatial structure
of the input beam. The existence of this coupling is of course well known,
and most of the time cavities are designed to avoid a degeneration between
a transverse mode and the fundamental mode of the cavity. Single mode
operation is thus achieved. However, as we will show in this paper, 
transverse modes have to be taken into account even for an input mode which is 
perfectly matched to the fundamental mode of a nondegenerate cavity. We will 
evaluate their effect on the quantum noise reduction and show that they can 
constitute a limiting factor for squeezing obtainable in the fundamental mode.

We study in particular the influence of higher order transverse cavity modes 
on optical bistabilty and on different kinds of quantum noise reduction.
The calculations are based on a linear input-output formalism where
fluctuations are linearized around the working point of the system \cite
{Reynaud2}. We describe the spatial beam structure in terms of
Gauss-Laguerre modes. The effect of the Kerr non-linearity is then
a coupling between the transverse modes. In section II, we derive the
coupling coefficients between the modes for a light beam propagating through
a thin Kerr medium. To present our approach we study in section III the 
propagation through a thin medium in free space. The modes
coupling is at the origin of non-linear losses for the fundamental mode. We
show that the fluctuations associated to these losses are not vacuum
fluctuations but correspond to phase conjugated noise. In section IV
we consider a Kerr medium inside an optical cavity. We derive analytic
expressions for the excess noise produced in the fundamental mode through
the coupling to higher order modes. In free space, this excess noise can not
be neglected since its strength is of the same order of magnitude than the
strength of the parametric noise modification. However, if the Kerr medium
is placed in a cavity, the cavity provides a mode selection which can lead
to an important reduction of the perturbing effect of higher order
transverses modes on the fundamental mode. When the cavity is non degenerate
and the transverse mode spacing is large compared to the cavity width, one
can perform a perturbative treatment in order to evaluate the influence of
transverse mode coupling on the fundamental cavity mode. The fundamental
mode may also accidentally coincide with a higher order mode corresponding
to an other longitudinal mode family. In this case, one can calculate steady
state and fluctuations by performing a two mode approximation described in
the second part of section IV. The analysis of transverse mode coupling will
finally enable us to estimate the validity limit of a single mode
approximation. The discussion of the results as well as some conclusive
remarks will be presented in section V. Details of the calculations are
exposed in the appendix.

\section{Nonlinear coupling of Gauss-Laguerre modes in a Kerr medium}

In this section, we study the propagation of a light beam of arbitrary
transverse shape through a thin Kerr medium. The analysis will supply us
with the coupling coefficients between different transverse modes due to
non-linear coupling. We consider a light beam of frequency $\omega _{{\rm L}}$
which is described by the complex amplitude $E\left( {\bf r}\right) $ of the
electric field where ${\bf r}$ stands for the three-dimensional position $%
\left( r,\theta ,x\right) $ in cylindrical coordinates: 
\begin{equation}
{\cal E}\left( {\bf r}\right) =e^{-i\omega _Lt}E\left( {\bf r}\right)
+e^{i\omega _Lt}E\left( {\bf r}\right) ^{\dagger }.  \label{FieldOperators}
\end{equation}
In the paraxial approximation, the propagation of the field amplitude along
the $x$ direction is given by 
\begin{equation}
\triangle _{_{\perp }}E\left( {\bf r}\right) +2ik_L\frac{\partial E\left( 
{\bf r}\right) }{\partial x}=-\frac{k_L^2}{\varepsilon _0}P_{{\rm NL}}\left( 
{\bf r}\right).  \label{paraxapp}
\end{equation}
$P_{{\rm NL}}\left( {\bf r}\right) $ is the non-linear polarization of the
medium, $\triangle _{_{\perp }}$ stands for the transverse Laplacian and $%
k_L=\omega _{{\rm L}}/c$ is the light wavevector. The electric field may be
expanded in Gauss-Laguerre modes $u_p^{\left( l\right) }\left( {\bf r}%
\right) $ which are solutions of the propagation equation in free space in
cylindrical coordinates: 
\begin{eqnarray}
E\left( {\bf r}\right) &=&\sum_{p,l}A_p^{\left( l\right) }\left( x\right)
u_p^{\left( l\right) }\left( {\bf r}\right), \\
u_p^{\left( l\right) }\left( {\bf r}\right) &=&\sqrt{\frac 2\pi }\sqrt{\frac{%
n!}{\left( n+l\right) !}}\frac 1{w\left( x\right) }\left( \sqrt{2}\frac r{%
w\left( x\right) }\right) ^l \nonumber \\ 
&&\times L_p^{\left( l\right) }\left( 2\frac{r^2}{%
w^2\left( x\right) }\right) \exp \left[ -\frac{r^2}{w^2\left( x\right) }%
-i\varphi _p^{\left( l\right) }\left( {\bf r}\right) \right].
\label{GaussianMode}
\end{eqnarray}
$L_p^{\left( l\right) }$ is the generalized Laguerre polynomial, $\varphi
_p^{\left( l\right) }\left( {\bf r}\right) $ is the phase of the mode, $%
\lambda $ is the laser wavelength, $w\left( x\right) $ the beam size, $w_0$
the beam waist and $l_R$ the Rayleigh divergence length: 
\begin{eqnarray}
\varphi _p^{\left( l\right) }\left( {\bf r}\right) &=&-\frac \pi \lambda 
\frac x{x^2+l_R^2}r^2+l\theta \nonumber \\
&&+\left( 2p+l+1\right) {\arctan }\left( \frac x{%
l_R}\right), \\
w^2\left( x\right) &=&w_0^2\left( 1+\frac{x^2}{l_R^2}\right), \\
l_R &=&\frac{\pi w_0^2}\lambda. \label{def1}
\end{eqnarray}
The fundamental Gaussian mode (TEM$_{00}$) corresponds to the indices $p=0$
and $l=0$. With the help of the normalization relations for Gauss-Laguerre
modes\cite{Abramovitz} one can derive the propagation equation for the mode
amplitudes $A_p^{\left( l\right) }\left( x\right) $ in presence of a
dielectric medium: 
\begin{equation}
\frac{\partial A_p^{\left( l\right) }\left( x\right) }{\partial x}=\frac 1{%
2ik_L}\int dr\ d\theta \ u_p^{\left( l\right) }\left( {\bf r}\right)
^{*}P\left( {\bf r}\right).  \label{AmpliProp}
\end{equation}

We will now assume the non-linearity to be represented by a lossless Kerr
medium. The polarization $P\left( {\bf r}\right) $ in this medium can be
expressed through the non-linear susceptibility: 
\begin{equation}
P_{{\rm NL}}\left( {\bf r}\right) =\varepsilon _0\chi ^{\left( 3\right)
}\left| E\left( {\bf r}\right) \right| ^2E\left( {\bf r}\right).  \label{PChi}
\end{equation}
From equation (\ref{AmpliProp}) we deduce the propagation equation for the
field amplitudes 
\begin{eqnarray}
\frac{\partial A_p^{\left( l\right) }\left( x\right) }{\partial x}&=&i\frac{%
k_L\chi ^{\left( 3\right) }}2\sum_{mno,qrs}\Lambda _{pqrs}^{\left(
lmno\right) }\left( x\right) A_q^{\left( m\right) }\left( x\right)
^{*}\nonumber \\ 
&&\times A_r^{\left( n\right) }\left( x\right) A_s^{\left( o\right) }\left(
x\right),  \label{KerrProp}
\end{eqnarray}
where we have introduced the coupling coefficients $\Lambda _{pqrs}^{\left(
lmno\right) }\left( x\right) $ between different modes: 
\begin{eqnarray}
\Lambda _{pqrs}^{\left( lmno\right) }\left( x\right) &=&\frac 1{\pi w^2}%
\lambda _{pqrs}^{\left( lmno\right) }\delta _{l+m,n+o}\nonumber \\
&&\times \exp \left[ -2i\left(
p+q-r-s\right) \arctan \left( \frac x{l_R}\right) \right],  \label{selrule} 
\end{eqnarray}
\begin{eqnarray}
\lambda _{pqrs}^{\left( lmno\right) } &=&2\int_0^\infty dv\
v^{l+m}L_p^{\left( l\right) }\left( v\right) L_q^{\left( m\right) }\left(
v\right) L_r^{\left( n\right) }\left( v\right) L_s^{\left( o\right) }\left(
v\right)\nonumber \\
&&\times \exp \left[ -2v\right].  \label{lambda2}
\end{eqnarray}
The dimensionless coefficients $\lambda _{pqrs}^{\left( lmno\right) }$ can
be evaluated with the use of generating functions. Details of the
calculation are exposed in appendix (\ref{app1}). In expression (\ref
{selrule}), one notices the presence of a selection rule $l+m=n+o$ which is
due to the revolution symmetry of the problem. When a circularly symmetric
beam ($l=0$) is sent into the medium, one can deduce from propagation
equation (\ref{KerrProp}) that it will only couple to higher order modes
having the same (circular) symmetry. In this case, we will drop the index $l$
\begin{eqnarray}
u_p\left( {\bf r}\right) &=&u_p^{\left( 0\right) }\left( {\bf r}\right), \\
\lambda _{pqrs} &=&\lambda _{pqrs}^{\left( 0000\right) }.  \label{up0}
\end{eqnarray}
In this discussion we have neglected the effect of transverse instabilities,
which arise in some region of parameter space\cite{Luigi2}, especially for
high intensities, which we do not want to treat here.

\section{Field propagation in free space}

We will now apply the treatment of non-linear coupling developed in the
previous section to the propagation of a fundamental Gaussian mode through a
Kerr medium. In order to simplify calculations we will consider a particular
situation which is however often encountered in experiments. We assume the
length $L$ of the medium to be much shorter than the Rayleigh divergence
length. If the medium is placed in the beam waist, equation (\ref{KerrProp})
for the propagation of the mode amplitudes simplifies to 
\begin{eqnarray}
\frac{\partial A_{p}\left( x\right) }{\partial x} &=&i\frac{\hat{K}}{L}%
\sum_{qrs}\lambda _{pqrs}A_{q}\left( x\right) ^{*}A_{r}\left( x\right)
A_{s}\left( x\right),  \label{KerrProp2} \\
\hat{K} &=&\frac{k\chi ^{\left( 3\right) }L}{2\pi w^{2}}.  \label{propsimpl}
\end{eqnarray}
Let the amplitude of the incoming beam be $A_{{\rm in}}$. We will now study
perturbatively the propagation of the fundamental mode inside the medium in
the limit of a small non-linearity, that is $\hat{K}\left| A_{{\rm in}%
}\right| ^{2}\ll 1$, condition which avoids the build-up of transverse
instabilities. The calculation will be performed up to second order in the
input intensity. Only in zero order the fundamental mode propagates
undisturbed. Solving propagation equation (\ref{KerrProp2}) in iterative
steps up to second order leads to the following expression for the
fundamental mode amplitude after propagation through the medium 
\begin{eqnarray}
A_{0}\left( L\right) &=&\left[ 1+i\varphi _{{\rm NL}}-\frac{1}{2}\varphi _{%
{\rm NL}}^{2}-\frac{1}{2}\gamma _{{\rm NL}}\right] A_{{\rm in}},  \label{A0}
\\
\varphi _{{\rm NL}} &=&\hat{K}\left| A_{{\rm in}}\right| ^{2}, \\
\gamma _{{\rm NL}} &=&\frac{1}{3}\varphi _{{\rm NL}}^{2}.  \label{FreeLoss}
\end{eqnarray}
In first order, the effect of the Kerr medium is a non-linear phase shift $%
\varphi _{{\rm NL}}$ proportional to the light intensity, like in the single
mode approximation. After isolating the second order contribution of the 
non-linear phase shift, one notices also a non-linear loss term 
$\gamma _{{\rm NL}}$ due to the energy transfer into higher order modes. It 
has to be noted that as soon as the non-linear phase shift becomes important, 
the associated losses can not be neglected and a perturbative treatment of the
coupling between the modes does not hold anymore.

Before calculating the propagation of fluctuations we will briefly introduce
our notation and the input-output formalism. A detailed presentation of this
method can be found in \cite{Courty}. We will denote $\alpha _{p}$ the
fluctuation amplitude of mode $p$ ($\alpha _{p}^{*}$ for its conjugate). To
simplify notations it is convenient to use a vectorial notation where
fluctuations of the light beam are completely described by a vector 
$\overrightarrow{\alpha}_{p}$ 
containing the two complex amplitudes: 
\begin{equation}
\overrightarrow{\alpha }_{p} =\left[ 
\begin{array}{c}
{\cal \alpha }_{p} \\ 
{\cal \alpha }_{p}^{*}
\end{array}
\right].  \label{alphap}
\end{equation}
We recall here that we will use a linear treatment of the fluctuations
around the working point of the system. All relations for the complex
amplitudes are therefore identical to the corresponding relations for the
annihilation and creation operators $a_{p}$ and $a_{p}^{\dagger }$ of mode $%
p $. Fluctuations are characterized by their correlation functions $%
C_{\alpha \beta }\left( \tau \right) $: 
\begin{equation}
C_{\alpha \beta }\left( \tau \right) =\left\langle \alpha \left( t+\tau
\right) \beta \left( t\right) \right\rangle  \label{correl}
\end{equation}
The brackets correspond to the quantum mechanical expectation value. To
calculate the explicit expressions for correlation functions it is
convenient to consider the problem in Fourier space. Transformations between
frequency and time domain are defined in the following way: 
\begin{equation}
f\left[ \omega \right] =\int dtf\left( t\right) e^{i\omega t}.
\label{Fourier}
\end{equation}
The Wiener Kinchin theorem establishes a relation between time and frequency
dependent correlation functions: 
\begin{eqnarray}
\left\langle \alpha \left[ \omega \right] \beta \left[ \omega ^{\prime
}\right] \right\rangle &=&2\pi \delta \left( \omega +\omega ^{\prime }\right)
C_{\alpha \beta }\left[ \omega \right] \nonumber \\
&=&2\pi \delta \left( \omega +\omega
^{\prime }\right) \int d\tau C_{\alpha \beta }\left( \tau \right) e^{i\omega
\tau }.  \label{Wiener}
\end{eqnarray}
Therefore fluctuations of a given mode may be completely described with the
help of the following $2\times 2$ matrix $\underline{V}_{p}$ which
contains all four field autocorrelation functions: 
\[
\underline{V}_{p}\left[ \omega \right]  =\left[ 
\begin{array}{cc}
C_{{\cal \alpha }_{p}{\cal \alpha }_{p}^{*}}\left[ \omega \right] & C_{{\cal %
\alpha }_{p}{\cal \alpha }_{p}}\left[ \omega \right] \\ 
C_{{\cal \alpha }_{p}^{*}{\cal \alpha }_{p}^{*}}\left[ \omega \right] & C_{%
{\cal \alpha }_{p}^{*}{\cal \alpha }_{p}}\left[ \omega \right]
\end{array}
\right]. 
\]

After this brief introduction we will now consider the propagation of
quantum fluctuations inside a Kerr medium. To this aim we will first derive
a propagation equation for fluctuations of mode $p$ using the propagation
equation for the mean field. Differentiating equation (\ref{KerrProp2})
leads to: 
\begin{equation}
\frac{\partial \alpha _p}{\partial x}=i\frac{\hat{K}}L\sum_{qrs}\lambda
_{pqrs}\left[ A_rA_s\alpha _q^{*}+2A_q^{*}A_r\alpha _s\right].
\end{equation}
In the case of a small non-linearity we develop the stationary mean field of
the fundamental mode up to second order and obtain a perturbative solution
for the modification $d\alpha _0$ of its fluctuations, which may be written
as the sum of a parametric term $\left. d\alpha _0\right| _{{\rm par}}$ and
additional fluctuations $\left. d\alpha _0\right| _{{\rm add}}$ coming from
higher order modes: 
\begin{eqnarray}
d\alpha _0 &=&\left. d\alpha _0\right| _{{\rm par}}+\left. d\alpha _0\right|
_{{\rm add}}, \\
\left. d\alpha _0\right| _{{\rm par}} &=&i\varphi _{{\rm NL}}\left( 2\alpha
_0+\alpha _0^{*}\right) -\frac{\varphi _{{\rm NL}}^2}3\left( 3\alpha
_0+2\alpha _0^{*}\right), \\
\left. d\alpha _0\right| _{{\rm add}} &=&2i\varphi _{{\rm NL}}\sum_q\lambda
_q\alpha _q+i\varphi _{{\rm NL}}\sum_q\lambda _q\alpha _q^{*} \nonumber \\
&&-\frac 32\varphi _{{\rm NL}}^2\sum_{qr}\lambda _q\lambda _{qr}\alpha
_r-\varphi _{{\rm NL}}^2\sum_{qr}\lambda _q\lambda _{qr}\alpha _r^{*}. 
\end{eqnarray}
For simplicity we have assumed the incoming field amplitude $A_{{\rm in}}$
to be real. The parametric term may be readily obtained by differentiating
equation (\ref{A0}). It corresponds to the usual parametric change of
quantum fluctuations and is obtained similarly in the single mode
approximation. The additional fluctuations can be understood as a
consequence of the losses experienced by the fundamental mode due to its
coupling to higher order modes. They correspond mainly to phase conjugated
noise. Indeed in first order, they may be written as: 
\begin{equation}
\left. d\alpha _0\right| _{{\rm add}}=i\varphi _{{\rm NL}}\sum_q\lambda
_q\left( 2\alpha _q+\alpha _q^{*}\right).
\end{equation}
Their correlation function 
\begin{equation}
\underline{V}_{{\rm add}} =\frac 13\hat{K}^2|A_{{\rm in}}|^4\left( 
\begin{array}{cc}
4 & -2 \\ 
-2 & 1
\end{array}
\right)  \label{Vaddfree}
\end{equation}
is characteristic of the mode coupling by the Kerr medium and we will find
it again in the following. Notice that even for small non-linearities this
additional noise is never negligible.

\section{Field propagation inside a bistable cavity}

We come now to the case where the non-linear medium is not in free space but
placed inside an optical cavity. We will consider the input field to be
perfectly matched to the fundamental Gaussian mode of the cavity 
($p=l=0$ for the input beam). Inside the cavity higher order transverse modes 
become
important due to non-linear coupling. However, because of the cylindrical
symmetry of the problem, we do not have to consider higher order modes with
non-circular symmetry ($l\neq 0$). The cavity is supposed to be single
ended, with an input mirror of amplitude transmissivity $\sqrt{2\gamma }$.
In the limit of a high finesse cavity it is possible to derive an evolution
equation for the intracavity field amplitude by evaluating the modification
of the field during one roundtrip. The roundtrip propagation of the
transverse mode $n,p$ gives rise to a phase shift $\varphi _{n,p}=\varphi
_{0}+n\varphi _{{\rm L}}+p\varphi _{{\rm T}}$ which is the sum of the
detuning $\varphi _{0}$ of the fundamental mode and integer multiple of the
longitudinal phase spacing $\varphi _{{\rm L}}$ and the transverse phase
spacing $\varphi _{{\rm T}}$ between different transverse modes. In a first
approach, we will suppose the spacing between longitudinal modes to be much
larger than $\varphi _{{\rm T}}$. In this limit, only one family of modes
corresponding a longitudinal index $n_{0}$ has to be considered. This
assumption will be given up in the last section, where we will consider also
the influence of different longitudinal cavity modes.

Compared to the case of field propagation through a non-linear medium placed
in free space, the field evolution equation inside the cavity has now to
contain additional terms which take into account losses of the intracavity
field as well as transmission of the input field through the cavity coupling
mirror: 
\begin{eqnarray}
\tau _c\frac{dA_p(t)}{dt}&=&\sqrt{2\gamma }A_p^{{\rm in}}(t)-\left( \gamma
+i\varphi _p\right) A_p(t)\nonumber \\
&&\times +i\hat{K}\sum_{qrs}\lambda
_{pqrs}A_q^{*}(t)A_r(t)A_s(t).  \label{EvMean}
\end{eqnarray}
$\tau _c$ is the roundtrip time of the field inside the cavity. In the last
term one recognizes the field modification due to the presence of the Kerr
medium. For an empty cavity at resonance, the mean amplitude of the
fundamental mode is $A_{\max }=\sqrt{2/\gamma }A^{{\rm in}}$ where the mean
amplitude of the fundamental mode is $A^{{\rm in}}=\left\langle A_0^{{\rm in}%
}\right\rangle $. We can then define normalized mode amplitudes ${\cal A}_p
=A_p/A_{\max }$, ${\cal A}_p^{{\rm in}} =A_p^{{\rm in}}/A^{{\rm in}}$ and $%
{\cal A}_p^{{\rm out}} =A_p^{{\rm out}}/A^{{\rm in}}$ respectively for the
intracavity, input and output fields. They obey a dimensionless evolution
equation 
\begin{equation}
\frac{d{\cal A}_p}{d\tau }={\cal A}_p^{in}-\left( 1+i\phi _p\right) {\cal A}%
_p+iK\sum_{qrs}\lambda _{pqrs}{\cal A}_q^{*}{\cal A}_r{\cal A}_s,
\label{EvCav}
\end{equation}
where $\tau =\gamma t/\tau _{{\rm c}}$ is a dimensionless time, $\phi
_p=\varphi _p/\gamma $ is the normalized phase shift of mode $p$ and $K=\hat{%
K}\left| A_{\max }\right| ^2/\gamma $ the normalized non-linear coupling
coefficient. The output field is given as a sum of the reflected input field
and the transmitted intracavity field: 
\begin{equation}
{\cal A}_p^{{\rm out}}(t)=-{\cal A}_p^{{\rm in}}(t)+2{\cal A}_p(t).
\label{Aout}
\end{equation}
In a single mode approximation, the intracavity intensity exhibits bistable
behaviour because of the intensity dependent phase shift. This phenomenon
occurs as soon as the non-linear phase shift $\phi _{{\rm NL}}$ of the field
is of the same order than the cavity width $\gamma $. Expressed with the
dimensionless non-linear coupling coefficient $K$ the corresponding condition
is $K\gtrsim 1$.

The evolution equation for the fluctuations is obtained by differentiating
equation (\ref{EvCav}) for the mode amplitudes around the stationary working
point 
\begin{eqnarray}
\partial _\tau {\cal \alpha }_p&=&{\cal \alpha }_p^{{\rm in}}-\left( 1+i\phi
_p\right) {\cal \alpha }_p\nonumber \\
&&+iK\sum_{qrs}\lambda _{pqrs}\left( {\bar{{\cal A}}}%
_r{\bar{{\cal A}}}_s{\cal \alpha }_q^{*}+2{\bar{{\cal A}}}_r^{*}{\bar{{\cal A%
}}}_s{\cal \alpha }_q\right).  \label{evfluct}
\end{eqnarray}
${\bar{{\cal A}}}_r$ is the stationary intracavity amplitude of mode $p$.
Although we consider an input field which is spatially matched to the
fundamental cavity mode, we nevertheless have to take into account
fluctuations entering the cavity through all possible input modes. The
output fluctuations can be deduced from 
\begin{equation}
{\cal \alpha }_p^{{\rm out}}=-{\cal \alpha }_p^{{\rm in}}+2{\cal \alpha }_p.
\label{alphaout}
\end{equation}
We have denoted here ${\cal \alpha }_p$, ${\cal \alpha }_p^{{\rm in}}$ and $%
{\cal \alpha }_p^{{\rm out}}$ the fluctuations of the normalized amplitudes $%
{\cal A}_p$, ${\cal A}_p^{{\rm in}}$ and ${\cal A}_p^{{\rm out}}$.

\subsection{Multimode perturbative approximation}

As mentioned in the beginning of this section we will first consider the
case where the transverse mode spacing is large compared to the cavity
bandwidth. ($\phi _{{\rm T}}=\varphi _{{\rm T}}/\gamma \gg 1$). Hence, when
the fundamental mode is nearly resonant, higher order transverse modes are
far off resonance and their amplitudes will be small compared to the
amplitude of the fundamental mode. This situation allows one to adopt a
perturbative treatment of the coupling between fundamental and higher order
modes, where the latter enter as first order quantities. Equation (\ref
{EvCav}) can now be solved in iterative steps up to second order in the
fundamental mode amplitude which leads to the new stationary state: 
\begin{eqnarray}
1&=&\left[ \left( 1+\mu _2\frac{K^2{\cal A}_0^4}{\phi _{{\rm T}}^2}\right)
+i\left( \phi _0-K{\cal A}_0^2-3\mu _1\frac{K^2{\cal A}_0^4}{\phi _T} \right. 
\right.\nonumber \\
&&\left.\left.+3\mu _2\frac{\phi _0K^2{\cal A}_0^4}{\phi _T^2}-15\mu _3\frac{K^3{\cal 
A}_0^6}{\phi
_{{\rm T}}^2}\right) \right] {\cal A}_0,  \label{Astat}
\end{eqnarray}
\begin{eqnarray}
\mu _1 &=&\sum_{p\neq 0}\frac{\lambda _p^2}p=\ln \left( \frac 43\right)
\simeq 0.288, \\
\mu _2 &=&\sum_{p\neq 0}\frac{\lambda _p^2}{p^2}\simeq 0.268, \\
\mu _3 &=&\sum_{p,q\neq 0}\frac{\lambda _p\lambda _q\lambda _{pq}}{pq}\simeq
0.197.  \label{mu}
\end{eqnarray}
$\phi _0$ is the normalized linear phase shift. The first bracket gives the
loss in the fundamental mode due to energy transfer to higher order modes.
The last three terms in the second bracket describe the non-linear phase
shift of the fundamental mode by the presence of higher order modes. Energy
transfer and phase shift arise as soon as higher order transverse modes
exist, but they are the smaller the more distant the perturbing transverse
modes are from a cavity resonance. However, they increase with increasing
intensity in the fundamental mode. At lowest order, their expressions are
given by 
\begin{eqnarray}
\phi _{{\rm NL}} &=&\frac{\varphi _{{\rm NL}}}\gamma =K|{\cal A}_0|^2, \\
\frac{\gamma _{{\rm NL}}}\gamma &=&\mu _2\frac{K^2|{\cal A}_0|^4}{\phi _{%
{\rm T}}^2}=\mu _2\frac{\phi _{{\rm NL}}^2}{\phi _{{\rm T}}^2}.  \label{loss}
\end{eqnarray}
By comparing this relation to the free space situation (cf. equation (\ref
{FreeLoss})), one notices that the
relations depend now on the cavity width $\gamma $ as well as on the
transverse mode spacing $\phi _{{\rm T}}$. In general the cavity may
therefore suppress or enhance the non-linear losses depending on whether the
fundamental mode and higher order modes are degenerate or not. For the non
degenerate cavity that we consider here the cavity reduces the non-linear
losses considerably.

We will now discuss the treatment of fluctuations where we will calculate
the field autocorrelation functions following the same iterative steps as in
the mean field calculation. Equation (\ref{evfluct}) couples fluctuations of
all transverse modes and their conjugates. In the frequency domain it takes
the following algebraic form 
\begin{eqnarray}
-i\omega \overrightarrow{\alpha}_{p}\left[ \omega \right] &=&\overrightarrow{\alpha 
}_{p}^{\rm 
in}\left[ \omega \right] -\left( 1+i\underline{\eta } \left( \phi _{0}+p\phi _{{\rm 
T}}-\underline{M}_{pp} \right)
\right) \overrightarrow{\alpha }_{p}\left[ \omega \right]  \nonumber \\
&&+i\underline{\eta} \sum_{q\neq p}\underline{M}_{pq} \overrightarrow{\alpha 
}_{q}\left[
\omega \right],  \label{fluctp}
\end{eqnarray}
where we have introduced the matrices $\underline{M}_{pq}$ and $\underline{\eta}$ 
defined by 
\begin{eqnarray}
\underline{\eta} &=&\left( 
\begin{array}{cc}
1 & 0 \\ 
0 & -1
\end{array}
\right), \\
\underline{M}_{pq} &=&K\sum_{rs}\lambda _{pqrs}\left( 
\begin{array}{cc}
2{\cal A}_{r}^{*}{\cal A}_{s} & {\cal A}_{r}{\cal A}_{s} \\ 
-{\cal A}_{r}^{*}{\cal A}_{s}^{*} & -2{\cal A}_{r}^{*}{\cal A}_{s}
\end{array}
\right).  \label{etaMpq}
\end{eqnarray}
$\omega $ is the fluctuation frequency normalized by the cavity bandwidth $%
\gamma $. By manipulating (\ref{fluctp}) the intracavity fluctuations 
$\overrightarrow{\alpha}_{p}$ of mode $p$ may be expressed as a
function of its incoming fluctuations $\overrightarrow{\alpha}_{p}^{\rm in}$ and of 
intracavity 
fluctuations $\overrightarrow{\alpha}_{q}$ in other modes: 
\begin{equation}
\overrightarrow{\alpha}_{p} =\underline{G}_{p}\left( \overrightarrow{\alpha}_{p}^{in} 
+i\underline{\eta} \sum_{q\neq p}\underline{M}_{pq} \overrightarrow{\alpha}_{q}\right).  
\label{propfluc}
\end{equation}
$\underline{G}_{p}$ is the propagator of an arbitrary mode $p\neq 0$
defined by 
\begin{equation}
\underline{G}_{p} =(1-i\omega +i\underline{\eta} \left( \phi _{0}+p\phi
_{{\rm T}}-\underline{M}_{pp} \right) )^{-1}.  \label{Gp}
\end{equation}
With the use of standard projector techniques, it is possible to express
fluctuations in the fundamental mode in an analogous way: 
\begin{equation}
\overrightarrow{\alpha}_{0} =\tilde{\underline{G}}_{0} \left( 
\overrightarrow{\alpha}_{0}^{{\rm 
in}} +i \underline{\eta} \sum_{q\neq
0}\tilde{\underline{T}}_{q} \overrightarrow{\alpha}_{q}^{\rm in}\right).  
\label{alpha0}
\end{equation}
Here we have taken explicitly into account the coupling of fluctuations in
the fundamental mode with input fluctuations in higher order modes by
introducing the transfer function $\tilde{\underline{T}}_{q} $. 
$\tilde{\underline{G}}_{0} $ is 
the propagator of the fundamental mode, which may
be written as 
\begin{equation}
\tilde{\underline{G}}_{0}  =(1-i\omega +i\underline{\eta} \phi
_{0}+\underline{R}_{0} )^{-1}.  \label{R01}
\end{equation}
$\underline{R}_{0}$ is a matrix corresponding to non-linear phase shifts
and losses. Equation (\ref{alpha0}) clearly shows that the intracavity
fluctuations of the fundamental mode are of the same order of magnitude than
its incoming fluctuations. In the case of large transverse mode spacing $\Phi _{{\rm 
T}}$ one can 
now analyze perturbatively the intracavity
fluctuations of higher order modes. We present the detailed calculation in
the appendix (\ref{app2}) where we give in particular the results for
non-linear phase shifts and losses as well as for the transfer function. Here
we will concentrate on the calculation and discussion of the physically more
interesting quantity, the fluctuations in the output beam.

We might first rewrite the intracavity fluctuations in a more compact form
as the sum of two contributions, one coming from the input fluctuations the
other from non-linear coupling: 
\begin{eqnarray}
\overrightarrow{\alpha }_0 &=&\tilde{\underline{G}}_0 \left( \overrightarrow{\alpha 
}_0^{{\rm 
in}} + \overrightarrow{\alpha }_{{\rm add}}\right), \nonumber \\
\overrightarrow{\alpha }_{{\rm add}} &=&i\underline{\eta} \sum_{q\neq
0}\tilde{\underline{T}}_q \overrightarrow{\alpha }_q^{{\rm in}}. 
\end{eqnarray}
The fluctuations in the output beam may now be calculated with the help of
expression (\ref{alphaout}): 
\begin{equation}
\overrightarrow{\alpha }_0^{{\rm out}}=\left( -1+2 \tilde{\underline{G}}_0 \right) 
\overrightarrow{\alpha }_0^{{\rm in}} +2i\tilde{\underline{G}}_0 \underline{\eta} 
\sum_{q\neq 
0}\tilde{\underline{T}}_q
\overrightarrow{\alpha }_q^{{\rm in}}.  \label{alpha0out}
\end{equation}
Their correlation function is found to split into the same distinct parts,
the input noise which is transformed by the cavity and the added noise due
to non-linear coupling: 
\begin{equation}
\underline{V}_0^{{\rm out}} =\left( -1+2\tilde{\underline{G}}_0 \right)
\underline{V}_{{\rm in}} \left( -1+2 \tilde{\underline{G}}_0\right)
^{\dagger }+4\tilde{\underline{G}}_0 \underline{V}_{{\rm add}} 
\tilde{\underline{G}}_0^{\dagger}. 
\label{Vout}
\end{equation}
The propagator $\tilde{\underline{G}}_0$ of fluctuations in the
fundamental mode is uniquely determined by the matrix $\underline{R}_0$ given in the
appendix (\ref{app2}). It is important to notice that all losses and phase
shifts due to coupling to higher order modes vanish at least with $1/\phi _{%
{\rm T}}$. $\underline{R}_0$ reduces therefore to a term corresponding to a simple
Kerr phase shift as soon as $\phi _{{\rm T}}$ becomes large. The incoming
beam is supposed to be in a coherent state so that its input fluctuations
correspond to vacuum noise. In this case the sum over all transverse modes
in (\ref{alpha0out}) may be performed and we find the correlation function
of the added noise to be 
\begin{equation}
\underline{V}_{{\rm add}} =\mu _2\frac{K^2{\cal A}_0^2}{\phi _{{\rm T}}^2}%
\left( 
\begin{array}{cc}
4\left| {\cal A}_0\right| ^2 & -2{\cal A}_0^2 \\ 
-2{\cal A}_0^{*2} & 1\left| {\cal A}_0\right| ^2
\end{array}
\right).  \label{Vadd}
\end{equation}
This result may be compared to the case of a fundamental Gaussian mode
propagating through a Kerr medium without a surrounding cavity (cf. equation
(\ref{Vaddfree})). Clearly, the additional noise is more and more suppressed
with an increasing spacing between transverse modes. Under these conditions
it is possible to describe the system in a single mode approximation.

\subsection{The two mode approximation}

In the previous sections, we have considered only one longitudinal mode and
assumed the corresponding fundamental mode and transverse modes to be non
degenerate. However, complications might arise when a transverse mode
corresponding to a different longitudinal mode coincides with the
fundamental mode of the cavity. In that case we expect important changes in
the mean intracavity intensity as well as in the quantum fluctuations of the
fundamental mode. We therefore consider now a situation where not only the
transverse modes (index $p$) of the cavity are important, but also its
longitudinal modes (index $n$) 
\begin{equation}
\varphi _{p}=\varphi _{0}+n\varphi _{{\rm L}}+p\varphi _{{\rm T}}.
\label{varphiL}
\end{equation}
$\varphi _{{\rm L}}$ is the free spectral range of the cavity. As before we
will assume in this section the fundamental mode and higher order transverse
modes of the same longitudinal mode family to be non degenerate, i.e. $%
\varphi _{{\rm T}}\gg \gamma $. However there is still the possibility of
degeneracy between two modes $a$ and $b$ when $n_{a}\varphi _{{\rm L}%
}+p_{a}\varphi _{{\rm T}}\simeq n_{b}\varphi _{{\rm L}}+p_{b}\varphi _{{\rm T%
}}$. In such a situation, the degeneracy concerns the whole family of modes.
Since the coupling coefficients between the fundamental mode and higher
order modes rapidly decrease with increasing mode order we will only
consider the first two modes of the family. 

The fundamental TEM$_{00}$ mode, denoted ${\cal A}$ with fluctuations $%
\alpha $, will be coupled to an arbitrary transverse mode ${\cal B}$ (with
fluctuations $\beta $) of order $p$. When the input field is perfectly matched
to the fundamental cavity mode, the evolution equations for the two modes, 
derived from equation (\ref{EvCav}), are 
\begin{eqnarray}
\partial _{\tau }{\cal A} &=&1-\left( 1+i\phi _{a}\right) {\cal A}+iK\left[ \left| 
{\cal 
A}\right| ^{2}{\cal A}\right.\nonumber \\
&& + \lambda _{p}\left( {\cal A}%
^{2}{\cal B}^{*}+2\left| {\cal A}\right| ^{2}{\cal B}\right) \nonumber \\
&& +\lambda_{pp}\left( {\cal B}^{2}{\cal A}^{*}+2\left| {\cal B}\right| ^{2}{\cal A}%
\right) +
\left. \lambda _{ppp0}\left| {\cal B}\right| ^{2}{\cal B}\right],  
\label{Abimode}  \\
\partial _{\tau }{\cal B} &=&0-\left( 1+i\phi _{b}\right) {\cal B}+iK\left[ \lambda 
_{p}\left| 
{\cal A}\right| ^{2}{\cal A}\right.\nonumber \\
&&+ \lambda_{pp}\left( {\cal A}^{2}{\cal B}^{*}+2\left| {\cal A}\right| ^{2}{\cal 
B}\right) 
\nonumber\\
&&\left. +\lambda _{ppp0}\left( {\cal B}^{2}{\cal A}^{*}+2\left| {\cal B}%
\right| ^{2}{\cal A}\right) +\lambda _{pppp}\left| {\cal B}\right| ^{2}{\cal B}\right]. 
\label{Bbimode}  
\end{eqnarray}
The coefficient $\lambda _{p}=1/2^{p}$ is at the origin of the energy
transfer from mode $a$ to mode $b$. Its value decreases rapidly with
increasing mode order $p$. This effect may be seen in figure 1 where we have
plotted the intracavity intensity of the fundamental mode as a function of
cavity detuning. Different bistability curves correspond to different
transverse mode orders $p$. The relative linear detuning between the
fundamental mode and the transverse mode is here chosen to be zero. As soon
as the mode order is larger than 4, the modification of the intracavity
intensity of the fundamental mode due to non-linear coupling with the
transverse mode becomes negligible compared to the single mode approximation
(dashed line). The influence of the non-linear coupling between the two modes
is expected to vary with their relative detuning. 
\begin{figure}
\centerline{\psfig{figure=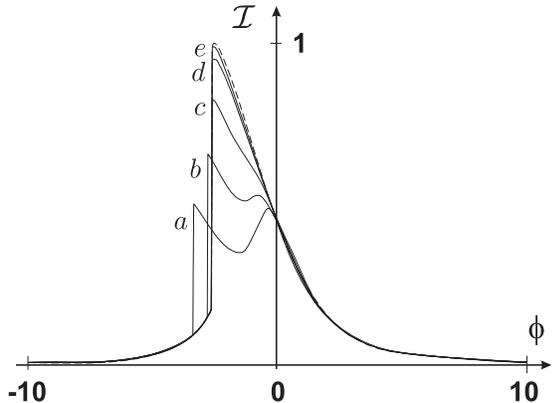,height=5.3cm}}
\caption{Bistability curves showing the intracavity intensity $%
{\cal I}=|{\cal A}|_0^2$ in the fundamental mode as a function of cavity
detuning $\phi $ in the two mode approximation. The non-linear coefficient is 
$K=2.5$, the relative detuning $\delta \phi =\phi _b-\phi _a=0$. Differents
curves in solid line correspond to increasing transverse mode order $p$: (a) 
$p=1$, (b) $p=2$, (c) $p=3$, (d) $p=4$, (e) $p=5$. The dashed line
corresponds to the single mode approximation.}
\end{figure}
\noindent The modification of the
bistability curve of the fundamental mode for various detunings between the
two modes can be seen in figure 2, where the fundamental mode is coupled to
the transverse mode $p=4$. Its perturbation is maximal when the transverse
mode is resonant with the fundamental mode. Since the two modes experience both linear 
and non-linear phase shifts, resonance occurs for a non-zero
linear detuning $\delta \phi =\phi _{b}-\phi _{a}$, where $\phi _{a}$ and $%
\phi _{b}$ denote respectively the linear phaseshifts of mode a and b.
\begin{figure}
\centerline{\psfig{figure=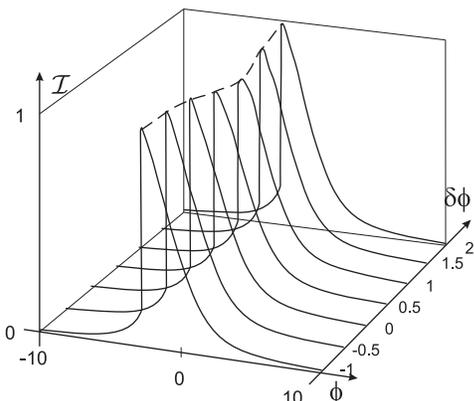,height=5.3cm}}
\caption{Intracavity intensity ${\cal I}=|{\cal A}|_0^2$ in the
fundamental mode as a function of cavity detuning $\phi $ in the two mode
approximation for a transverse mode $p=4$. The non-linear coefficient is $%
K=2.5$. Different curves correspond to different detunings $\delta \phi $
between the two modes. The dashed line marks the maxima of the bistability
curves.}
\end{figure}

We will now analyze the influence of the non-linear coupling on fluctuations
of the fundamental mode. It is first worth noting that the behaviour of
coefficient $\lambda _{pp}$ for a large transverse mode order is 
\begin{equation}
\lambda _{pp}=\frac{1}{4^{p}}\frac{\left( 2p\right) !}{\left( p!\right) ^{2}}%
\sim \sqrt{\frac{1}{p\pi }}.  \label{nu2}
\end{equation}
This expression looks similar to a cross Kerr effect between the two modes.
The value of this coefficient is decreasing only slowly with increasing mode
order. The consequences will be discussed in the following paragraph.

The equation for the evolution of fluctuations in the two modes is obtained
by linearizing the mean field equations (\ref{Abimode},\ref{Bbimode}). It is
convenient to represent these fluctuations in a four element vector 
$\overrightarrow{\alpha}$ 
\begin{equation}
\overrightarrow{\alpha} =\left[ 
\begin{array}{c}
{\cal \alpha } \\ 
{\cal \alpha }^{*} \\ 
{\cal \beta } \\ 
{\cal \beta }^{*}
\end{array}
\right],  \label{alphabi}
\end{equation}
which satisfies the evolution equation: 
\begin{equation}
\partial _\tau \overrightarrow{\alpha } =\overrightarrow{\alpha}^{{\rm in}%
}-\left( 1+i\underline{\Phi} +iK\underline{M}\right) \overrightarrow{\alpha}.  
\label{alphabiev}
\end{equation}
$\underline{\Phi}$ is a diagonal four by four matrix containing the
linear detunings $\phi _a$ and $\phi _b$ for the two modes. In the matrix 
$\underline{M}$ we have 
collected all non-linear terms. The exact
expressions of both $\underline{\Phi}$ and $\underline{M}$ can be found in appendix 
(\ref{app3}%
). The input-output relations for field fluctuations in terms of correlation
matrices $V_{{\rm out}}$ and $V_{{\rm in}}$ lead to the following relation
between the correlation functions 
\begin{equation}
\underline{V}^{{\rm out}} =\left( \frac{1-\underline{M}+i\omega }{%
1+\underline{M} -i\omega }\right) \underline{V}^{{\rm in}} \left( \frac{%
1-\underline{M} +i\omega }{1+\underline{M} -i\omega }\right) ^{\dagger }.
\label{Voutbi}
\end{equation}
Again the input field is supposed to be in the vacuum state. With the help
of the exact expression for $\underline{M}$ (cf. equation (\ref{M})) it is then
possible to calculate the correlation functions of the output field. We will
discuss the result by showing the corresponding numerical noise spectra.

Figure 3 shows the optimum noise reduction in the fundamental mode as a
function of the dimensionless frequency $\omega $. We have supposed
fluctuations to be measured in a homodyne detection\cite{Caves,Yuen,Shapiro}
where the output beam is superimposed with a local oscillator whose spatial
structure corresponds to the TEM$_{00}$ mode. 
The shot noise level is normalized to 0, perfect squeezing corresponds to $-1$. 
For all curves, the
working point has been chosen on the upper branch of the bistability curve
near the bistability turning point. Different curves correspond to transverse mode 
orders. Clearly, the effect of different non-linear coupling compared
to the single mode approximation (dashed line) is a reduction and even
suppression of the squeezing in the fundamental mode due to added 
\begin{figure}
\centerline{\psfig{figure=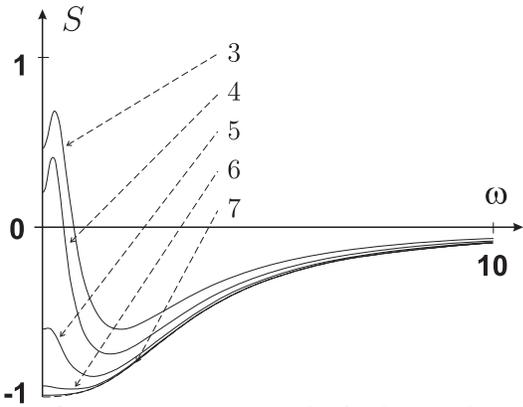,height=5.3cm}}
\caption{Optimum squeezing in the fundamental mode versus normalized
frequency $\omega $ for a coupling to various higher order modes. The local
oscillator is supposed to be matched to the TEM$_{00}$ mode. The working
point is on the upper branch of the bistability curve close to the turning
point. The non-linear coefficient is $K=2.5$, the relative detuning $\delta
\phi =1$. Different curves in solid line correspond to increasing transverse
mode order $p$ as mentioned in the figure. The dashed line corresponds to
the single mode approximation.}
\end{figure}
\noindent 
noise
coming from the perturbing mode, even though the spatial structure of the
fundamental mode is perfectly matched to a cavity mode. It has to be noticed
that this perturbation is higher than what would have been expected from the
examination of the bistability curves: for the mode $p=5$, the steady state
(figure 1(e)) is almost the steady state of the single mode cavity and only
a small fraction of the incoming photons is transferred to the second mode.
However, the added noise is sufficient to degrade significantly the
squeezing around zero frequency. This large modification is due to the non
negligible value of the ''cross Kerr'' coupling $\lambda _{pp}$ between the
two modes, which is still important for $p=5$.

The influence of the relative detuning between the two modes is shown in
figure 4 where the squeezing spectra of the fundamental mode are plotted for
a coupling with the transverse mode $p=4$ and various relative detunings. A
comparison with figure 2 shows that fluctuations are much more sensitive
to the detuning between the two modes than the mean field. In particular,
quadrature squeezing in the fundamental mode is found to be completely
suppressed in some range of cavity detuning. This shows that even in a
perfectly matched cavity the non-linear transverse mode coupling can limit
the value of quadrature squeezing attainable.

However, it is possible to recover an optimal quantum noise reduction in the
output beam, if the spatial structure of the local oscillator is chosen more
carefully. To illustrate this effect we have plotted in figure 5 the optimum
squeezing in a field quadrature corresponding to a linear superposition
between the fundamental and the pertubing mode. Fundamental and higher order
mode are supposed to be resonant so that the perturbation due to the mode
coupling is maximal. Even for a coupling with a mode of an order as low as $%
p=3$ the squeezing at
\begin{figure}[h]
\centerline{\psfig{figure=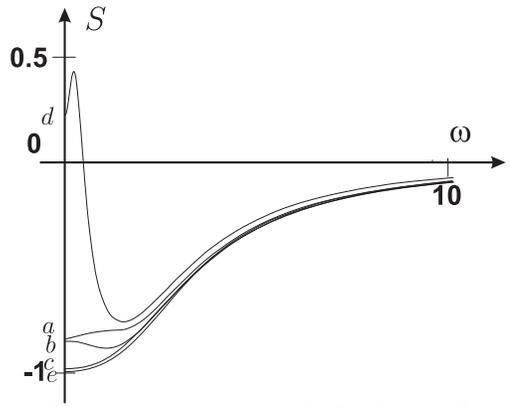,height=5.3cm}}
\caption{Optimum squeezing in the fundamental mode versus normalized
frequency $\omega $ for various relative detunings $\delta \phi $. The
working point is on the upper branch of the bistability curve close to the
turning point. The non-linear coefficient is $K=2.5$. Different curves in
solid line correspond to: (a) $\delta \phi =2$, (b) $\delta \phi =1$, (c) $%
\delta \phi =0$, (d) $\delta \phi =-1$, (e) $\delta \phi =-2$.}
\end{figure}

\noindent zero frequency is here almost perfect. This result has
to be seen in comparison with figure 3 where we had shown the optimum
squeezing in the output beam for the same experimental parameters but
measured with a local oscillator the spatial structure of which was not
optimized. It can also be noticed that here the noise reductions obtained
for a coupling with different transverse modes show little differences. This
can be understood from the squeezing mechanism. The bistability turning
point is a critical point at which fluctuations in one of the field
quadratures diverge. Since the system is lossless, fluctuations in the
conjugated field quadrature have to tend to zero when the turning point is
reached.

Compared to quadrature squeezing the influence of non-linear coupling on
intensity squeezing is very differ-
\begin{figure}
\centerline{\psfig{figure=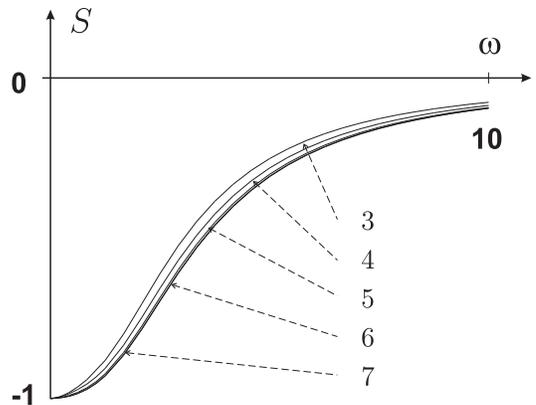,height=5.3cm}}
\caption{Optimum squeezing versus normalized frequency $\omega $ for a
coupling to various higher order modes. The local oscillator is supposed to
have an optimzed spatial structure. The working point is on the upper branch
of the bistability curve close to the turning point. The non-linear
coefficient is $K=2.5$, the relative detuning $\delta \phi =1$. Different
curves in solid line correspond to increasing transverse mode order $p$ as
mentioned in the figure. The single mode approximation coincides with the
curve of mode 7.}
\end{figure}

\noindent ent. Figure 6 shows a comparison of the
attainable intensity squeezing for a coupling to various modes. Clearly,
this squeezing is very robust against the mode coupling. Furthermore the
curves for various modes show little differences. 
\begin{figure}
\centerline{\psfig{figure=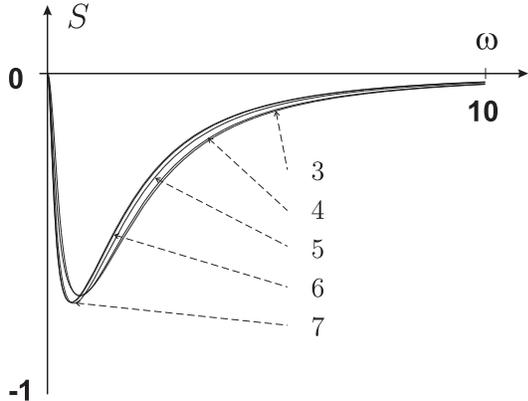,height=5.3cm}}
\caption{Intensity squeezing versus normalized frequency $\omega $ for a
coupling to various higher order modes. The working point is on the upper
branch of the bistability curve close to the turning point. The non-linear
coefficient is $K=2.5$, the relative detuning $\delta \phi =1$. Different
curves in solid line correspond to increasing transverse mode order $p$. The
single mode approximation coincides with the curve $p=7$.}
\end{figure}
\noindent As expected from the
squeezing by a lossless medium the noise at zero frequency is not modified
since the total photon number is conserved. It also has to be noticed that
for a coupling with mode $p=3$, the energy transfer from the fundamental
mode to the transverse mode is not negligible. As a consequence the spatial
structure of the beam going out of the cavity is quite different from the
fundamental mode. Its intensity fluctuations are nevertheless reduced and
this reduction corresponds to the one that could be expected from the single
mode situation.

\section{Conclusion}

In this paper, we have studied the non-linear coupling between transverse
modes in a Kerr medium and its effect on bistability and quantum noise if
the medium is placed inside a cavity. Whereas in free space the non-linear
coupling becomes important as soon as the non-linear phase shift is
appreciable, the use of an optical cavity can reduce significantly the
perturbing effect of the transverse modes by selecting only a few modes.
When fundamental and higher order modes belong to the same longitudinal mode
family and only the fundamental mode is resonant, it is possible to treat
higher order transverse modes perturbatively. The non-linear coupling leads
to losses for the fundamental mode associated with an excess noise which
contains a contribution from phase conjugated noise. The non-linear loss
scales with the inverse of transverse mode spacing. Perturbations due to
higher order transverse modes can therefore be neglected as soon as the
transverse mode spacing becomes large compared to the cavity width. In this
case a single mode approximation is valid.

In an experiment the fundamental mode may accidentally coincide with a
transverse mode corresponding to a different longitudinal mode family. In
this case, one can perform the calculations in a two mode approximation. We
have shown that even if the spatial structure of the input beam is perfectly
adapted to the fundamental mode of the cavity, the non-linear coupling
produces an excess noise in the fluctuations of the fundamental mode and
therefore can limit the optimum squeezing. The perturbation always arises
when low order modes ($p\leq 6$) coincide resonantly with the fundamental
mode during the resonance scan of the cavity. The reduction or even suppression
of the obtainable squeezing occurs
then in the central part of the noise spectrum. Quadrature squeezing in the
fundamental mode is thus sensitive on perturbations by higher order modes.
However, it is possible to recover the optimum squeezing in the output beam
if the spatial structure of the local oscillator is optimized. Fluctuations
are then not measured in the fundamental mode but in a mode corresponding to
a linear superposition of the fundamental and the perturbing mode.

In contrast to quadrature squeezing quantum noise reduction in the total
output intensity of the fundamental mode turns out to be very robust to
perturbations by higher order modes. Hence, a Kerr medium inside an optical
cavity remains a good intensity ``noise eater'' even when taking into
account higher order transverse modes.

{\bf Acknowledgments:} Thanks are due to S. Reynaud, C. Fabre, C. Schwob and
L. Hilico for discussions.

\appendix 

\section{Nonlinear coupling coefficients}

\label{app1}

The calculation of expression for the non-linear coupling coefficients $\lambda 
_{pqrs}^{\left( 
lmno\right) }$ between Gauss-Laguerre modes in a
Kerr medium can be performed by making use of the generating functions of
the Laguerre Polynomials $G_n^{\left( l\right) }\left( x,z\right) $: 
\begin{eqnarray}
L_n^{\left( l\right) }\left( x\right) &=&\frac 1{n!}\left. \frac{\partial ^n%
}{\partial z^n}\right| _{z=0} \\
G_n^{\left( l\right) }\left( x,z\right) &=&\frac 1{\left( 1-z\right) ^{l+1}}%
\exp \left( \frac{xz}{z-1}\right).
\end{eqnarray}
The coupling coefficient may then be reexpressed as a function of the
generating functions 
\begin{eqnarray}
\lambda _{pqrs}^{\left( lmno\right) }&=& \frac {2}{p!q!r!s!} 
\frac{\partial ^{p+q+r+s}}{\partial z_1^p\partial z_2^q\partial
z_3^r\partial z_4^s}  
\int_0^\infty dx x^{l+m}G_p^{\left( l\right) }\left(x,z_1\right) 
\nonumber \\
&&\times G_q^{\left( l\right) }\left( x,z_2\right) G_r^{\left( l\right)}\left( 
x,z_3\right) 
G_s^{\left( l\right) }\left( x,z_4\right) \nonumber \\
&&\times \exp \left[-2x\right] \left. \right| _{z_1=z_2=z_3=z_4=0}. 
\end{eqnarray}
In cylindrical symmetry and with $r=s=0$ they simplify to 
\begin{eqnarray}
\lambda _{pq=} &=&\frac 2{p!q!}\frac{\partial ^{p+q}}{\partial
z_1^p\partial z_2^q}\int_0^\infty dx \frac 1{\left( 1-z_1\right) \left(
1-z_2\right) ^1}  \\
&&\times \exp x \left( \frac{z_1}{z_1-1}+\frac{z_2}{z_2-1}-2\right) \left.
\right| _{z_1=z_2=0} \nonumber \\
&=& \frac 2{p!q!}\frac{\partial ^{p+q}}{\partial z_1^p\partial z_1^q}%
\frac 1{2-z_1-z_2}\left. \right| _{z_1=z_2=0}=\frac 1{2^{p+q}}\frac{\left(
p+q\right) !}{p!q!} \nonumber
\end{eqnarray}

\section{The multimode approximation}

\label{app2}

The evolution equation of mode p may be rewritten as 
\begin{equation}
\overrightarrow{\alpha }_p =\underline{G}_p \overrightarrow{\alpha }_p^{%
{\rm in}} +i\underline{G}_p \underline{\eta}  \underline{M}_{p0}\overrightarrow{\alpha 
}_0 
+i\underline{G}_p \underline{\eta}
 \sum_{q\neq p,0}\underline{M}_{pq} \overrightarrow{\alpha }_q.
\label{evfond}
\end{equation}
By iterating once and inserting the result in the evolution equation of the
fundamental mode (\ref{evfond}) gives 
\begin{eqnarray}
\underline{G}_0 ^{-1}\overrightarrow{\alpha }_0 &=&\overrightarrow{\alpha 
}_0^{{\rm in}} -\sum_{q\neq 0}\underline{ \eta } \underline{M}_{0q} \underline{G}_q 
\underline{ 
\eta } \underline{M}_{q0}
\overrightarrow{\alpha }_0 \nonumber \\
&-&i\sum_{q\neq 0}\sum_{r\neq q,r\neq 0}\underline{ \eta } \underline{M}_{0q} 
\underline{ G}_q 
\underline{ \eta } \underline{ M}_{qr}
\underline{ G}_r \underline{ \eta } \underline{ M}_{r0} \overrightarrow{%
\alpha }_0  \nonumber \\
&&+i\underline{ \eta } \sum_{q\neq 0}\underline{ M}_{0q} \underline{ G}_q
\overrightarrow{\alpha }_q^{{\rm in}} \nonumber \\
&&-\sum_{q\neq 0}\sum_{r\neq
q,0}\underline{ \eta } \underline{ M}_{0q} \underline{ G}_q \underline{ \eta
} \underline{ M}_{qr} \underline{ G}_r \overrightarrow{\alpha }_r^{{\rm %
in}}.
\end{eqnarray}
This equation may be resolved for $\alpha _0$, and fluctuations in the
fundamental mode may now be rewritten in the more convenient form 
\begin{equation}
\overrightarrow{\alpha }_0 =\underline{ \tilde{G}}_0 \left( \overrightarrow{\alpha 
}_0^{{\rm in}} 
+i\underline{ \eta } \sum_{q\neq
0}\underline{ \tilde{T}}_q \overrightarrow{\alpha }_q^{{\rm in}} \right),
\end{equation}
with the propagator $[\tilde{G}_0]$ for the fundamental mode and the
transfer matrix $\underline{ \tilde{T}}_q $ 
\begin{eqnarray}
\underline{ \tilde{G}}_0 &=&\left( 1-i\omega +\underline{ R}_0 +i\underline{
\eta } \phi _0\right) ^{-1}, \\
\underline{ \tilde{T}}_q &=&\underline{ M}_{0q} \underline{ G}_q
+i\sum_{r\neq q,0}\underline{ M}_{0q} \underline{ G}_q \underline{ \eta }
\underline{ M}_{qr} \underline{ G}_r, \\
\underline{ R}_0 &=&-i\underline{ \eta}  \underline{ M}_{00} +\sum_{q\neq
0}\underline{ \eta } \underline{ M}_{0q} \underline{ G}_q \underline{ \eta
} \underline{ M}_{q0} \\
&+&i\sum_{q\neq 0}\sum_{r\neq q,0}b\underline{ \eta } \underline{ M}_{0q}
\underline{ G}_q \underline{ \eta } \underline{ M}_{qr} \underline{ G}_r
\underline{ \eta } \underline{ M}_{r0} \nonumber .
\end{eqnarray}
In the multimode perturbative expansion the expression of $\underline{ R}_0$
can be simplified to 
\begin{eqnarray}
\underline{ R}_0 &=&+\mu _2\frac{K^2{\cal A}_0^2}{\phi _{{\rm T}}^2}\left( 
\begin{array}{cc}
3\left| {\cal A}_0\right| ^2 & 2{\cal A}_0^2 \\ 
2{\cal A}_0^{*2} & 3\left| {\cal A}_0\right| ^2
\end{array}
\right) -3i\mu _2\frac{\omega K^2{\cal A}_0^4}{\phi _{{\rm T}}^2} \nonumber \\
&&-iK\left( 
\begin{array}{cc}
2\left| {\cal A}_0\right| ^2 & {\cal A}_0^2 \\ 
-{\cal A}_0^{*2} & -2\left| {\cal A}_0\right| ^2
\end{array}
\right) -3i\frac{K^2{\cal A}_0^2}{\phi _{{\rm T}}}\left( \mu _1-\mu _2\frac{%
\phi _0}{\phi _{{\rm T}}}\right) \nonumber \\
&&\times \left( 
\begin{array}{cc}
3\left| {\cal A}_0\right| ^2 & 2{\cal A}_0^2 \\ 
-2{\cal A}_0^{*2} & -3\left| {\cal A}_0\right| ^2
\end{array}
\right) \nonumber \\
&&-12i\mu _3\frac{K^3{\cal A}_0^4}{\phi _{{\rm T}}^2}\left( 
\begin{array}{cc}
4\left| {\cal A}_0\right| ^2 & 3{\cal A}_0^2 \\ 
-3{\cal A}_0^{*2} & -4\left| {\cal A}_0\right| ^2
\end{array}
\right).  \label{R0}
\end{eqnarray}
The specific expression for the transfer matrix evaluates to 
\begin{eqnarray}
\underline{ \tilde{T}}_q &=&\left( -\frac{i\lambda _qK}{q\phi _T}+i\frac{%
\lambda _q\phi _0K}{q^2\phi _T^2}-2i\frac{K^2{\cal I}_0}{\phi _T^2}%
\sum_{r\neq 0}\frac{\lambda _r\lambda _{qr}}{qr}\right) \nonumber \\
&&\times \left( 
\begin{array}{cc}
2\left| {\cal A}_0\right| ^2 & {\cal A}_0^2 \\ 
-{\cal A}_0^{*2} & -2\left| {\cal A}_0\right| ^2
\end{array}
\right) \nonumber \\
&&-i\frac{K^2}{\phi _T^2}\sum_{r\neq 0}\frac{\lambda _q\lambda _{qr}}{qr}%
\left( 
\begin{array}{cc}
5\left| {\cal A}_0\right| ^2 & 4{\cal A}_0^2 \\ 
-4{\cal A}_0^{*2} & -5\left| {\cal A}_0\right| ^2
\end{array}
\right) \nonumber \\
&&+\frac{\lambda _qK}{q^2\phi _T^2}\left( 1-i\omega \right) \left( 
\begin{array}{cc}
2\left| {\cal A}_0\right| ^2 & {\cal A}_0^2 \\ 
{\cal A}_0^{*2} & 2\left| {\cal A}_0\right| ^2
\end{array}
\right).  \label{Tq}
\end{eqnarray}

\section{The two mode approximation}

\label{app3}

The evolution equation for the complex amplitude of the two modes is 
\begin{equation}
\partial _\tau \overrightarrow{\alpha } =\overrightarrow{\alpha }^{{\rm in}} -\left( 
1+i\underline{ \Phi}  +iK\underline{ M} \right) \overrightarrow{\alpha },
\end{equation}
with the four dimensional matrices $[\phi ]$ and $[M]$: 
\begin{equation}
\begin{array}{ccc}
\underline{ \Phi}  =\left[ 
\begin{array}{cc}
\Phi _a & 0 \\ 
0 & \Phi _b
\end{array}
\right], & \Phi _a=\left[ 
\begin{array}{cc}
\phi _a & 0 \\ 
0 & -\phi _a
\end{array}
\right], & \Phi _b=\left[ 
\begin{array}{cc}
\phi _b & 0 \\ 
0 & -\phi _b
\end{array}
\right],
\end{array}
\end{equation}
\begin{equation}
\underline{ M} =\underline{ M}_0 +\lambda _p\underline{ M}_1 +\lambda
_{pp}\underline{ M}_2 +\lambda _{ppp0}\underline{ M}_3 +\lambda
_{pppp}\underline{ M}_4,  \label{M}
\end{equation}
\begin{eqnarray}
&& 
\begin{array}{cc}
M_a=\left[ 
\begin{array}{cc}
-2\left| {\cal A}\right| ^2 & -{\cal A}^2 \\ 
{\cal A}^{*2} & 2\left| {\cal A}\right| ^2
\end{array}
\right], & M_b=\left[ 
\begin{array}{cc}
-2\left| {\cal B}\right| ^2 & -{\cal B}^2 \\ 
{\cal B}^{*2} & 2\left| {\cal B}\right| ^2
\end{array}
\right]
\end{array}
\nonumber \\[5mm]
&& 
\begin{array}{cc}
M_{ab}=\left[ 
\begin{array}{cc}
-2\left( {\cal A^{*}B+AB^{*}}\right) & -2{\cal AB} \\ 
2{\cal AB} & 2\left( {\cal A^{*}B+AB^{*}}\right)
\end{array}
\right] & 
\end{array}
\end{eqnarray}
\begin{equation}
\begin{array}{cc}
\underline{ M}_0 =\left[ 
\begin{array}{cc}
M_a & 0 \\ 
0 & 0
\end{array}
\right], & \underline{ M}_1 =\left[ 
\begin{array}{cc}
M_{ab} & M_a \\ 
M_a & 0
\end{array}
\right], \nonumber \\[5mm]
\underline{ M}_2 =\left[ 
\begin{array}{cc}
M_b & M_{ab} \\ 
M_{ab} & M_a
\end{array}
\right], 
& \underline{ M}_3 =\left[ 
\begin{array}{cc}
0 & M_b \\ 
M_b & M_{ab}
\end{array}
\right], \nonumber \\[5mm]
\underline{ M}_4 =\left[ 
\begin{array}{cc}
0 & 0 \\ 
0 & M_b
\end{array}
\right]. 
\end{array}
\end{equation}

\end{document}